\documentclass{emulateapj}
\usepackage{apjfonts}

 \newcommand{\eg}{\emph{e.g.}}
 \newcommand{\kms}{\mbox{km\ \ensuremath{\rm{s}^{-1}}}}
 \newcommand{\cold}[2]{$#1\times10^{#2}$ cm$^{-2}$}
\newcommand{\be}{\begin{equation}}
\newcommand{\ee}{\end{equation}}

\usepackage{graphicx}	% Including figure files

\begin{document}

\title{Deep K-band observations of TMC-1 with the Green Bank Telescope: Detection of HC$_7$O,  non-detection of HC$_{11}$N, and a search for new organic molecules}

\author{M. A. Cordiner\altaffilmark{1,2}, S. B. Charnley\altaffilmark{1}, Z. Kisiel\altaffilmark{3}, B. A. McGuire\altaffilmark{4},  Y.-J. Kuan\altaffilmark{5,6}}

\altaffiltext{1}{Astrochemistry Laboratory and the Goddard Center for Astrobiology, NASA Goddard Space Flight Center, 8800 Greenbelt Road, MD 20771, USA.}
\email{martin.cordiner@nasa.gov}
\altaffiltext{2}{Institute for Astrophysics and Computational Sciences, The Catholic University of America, Washington, DC 20064, USA.}
\altaffiltext{3}{Institute of Physics, Polish Academy of Sciences, Al. Lotnikow 32/46, 02-668 Warszawa, Poland.}
\altaffiltext{4}{National Radio Astronomy Observatory, Charlottesville, VA 22903, USA.}
\altaffiltext{5}{National Taiwan Normal University, Taipei 116, Taiwan, ROC.}
\altaffiltext{6}{Institute of Astronomy and Astrophysics, Academia Sinica, Taipei 106, Taiwan, ROC.}

% Abstract of the paper
\begin{abstract}
The 100 m Robert C. Byrd Green Bank Telescope K-band (KFPA) receiver was used to perform a high-sensitivity search for rotational emission lines from complex organic molecules in the cold interstellar medium towards TMC-1 (cyanopolyyne peak), focussing on the identification of new carbon-chain-bearing species as well as molecules of possible prebiotic relevance. 
We report a detection of the carbon-chain oxide species HC$_7$O and derive a column density of $(7.8\pm0.9)\times10^{11}$~cm$^{-2}$. This species is theorized to form as a result of associative electron detachment reactions between oxygen atoms and C$_7$H$^-$, {and/or reaction of C$_6$H$_2$$^+$ with CO (followed by dissociative electron recombination)}. Upper limits are given for the related HC$_6$O, C$_6$O and C$_7$O molecules. 
In addition, we obtained the first detections of emission from individual $^{13}$C isotopologues of HC$_7$N, and derive abundance ratios HC$_7$N/HCCC$^{13}$CCCCN = $110\pm16$ and HC$_7$N/HCCCC$^{13}$CCCN = $96\pm 11$, indicative of significant $^{13}$C depletion in this species relative to the local interstellar elemental $^{12}$C/$^{13}$C ratio of 60-70. 
 The observed spectral region covered two transitions of HC$_{11}$N, but emission from this species was not detected, and the corresponding column density upper limit is \cold{7.4}{10} (at 95\% confidence). This is significantly lower than the value of \cold{2.8}{11} previously claimed by \citet{bel97} and confirms the recent non-detection of HC$_{11}$N in TMC-1 by \citet{loo16}.
 Upper limits were also obtained for the column densities of malononitrile and the  nitrogen heterocycles quinoline, isoquinoline and pyrimidine.

\end{abstract}


%%%%%%%%%%%%%%%%%%%%%%%%%%%%%%%%%%%%%%%%%%%%%%%%%%

%%%%%%%%%%%%%%%%% BODY OF PAPER %%%%%%%%%%%%%%%%%%

\section{Introduction}

Radio spectroscopy is a powerful and rigorous technique for the detection of new molecules in the dense interstellar medium.  
Organic molecules are commonly observed in many different astronomical sources \citep{her09} and the  fact that many of the known interstellar organics  are also present in protoplanetary disks and  in comets \citep[\eg][]{obe15,wal16,coc15,biv15,alt16} supports the view that  interstellar clouds could plausibly be the first formation sites for  the prebiotic molecules that may have been delivered to the  early Earth by comets \citep[\eg][]{mum11}. 

The dark interstellar cloud TMC-1 has proven to be the archetype for studies of cold cloud organic chemistry \citep{kai04,sak13,mce13}.
Observations show that the inventory of TMC-1 is dominated by long carbon chain molecules:
the cyanopolyynes (HC$_{2n+1}$N, $n=1,2,3,4$; CH$_3$C$_3$N, CH$_3$C$_5$N), carbon chain radicals  (C$_2$H, C$_3$H, C$_4$H, C$_5$H, C$_6$H, C$_8$H,  C$_2$N, C$_3$N, C$_5$N) and related anions, linear  carbenes (H$_2$CCC, H$_2$CCCC, H$_2$CCCCCC),
polyacetylenic molecules (CH$_3$C$_2$H, CH$_3$C$_4$H) and smaller ring molecules such as $c$-C$_3$H and $c$-C$_3$H$_2$ \citep{gra16,ohi16}.

  Cyanodecapentayne (HC$_{11}$N) has long been considered as the largest gas-phase molecule identified in dense interstellar clouds. The first reported measurement of HC$_{11}$N in TMC-1 was by \citet{bel97}, who detected two apparent transitions ($J=39-38$ and $J=38-37$) in the Ku band using the NRAO 140-foot telescope, from which they derived a column density of \cold{2.8}{11} (assuming a rotational temperature of 10~K). However, there have since been no reported confirmations of HC$_{11}$N in the literature, either in TMC-1 or any other astronomical object.  \citet{loo16} searched for and failed to detect six transitions of HC$_{11}$N in the Ku band in TMC-1 using the Green Bank Telescope (GBT), including the $J=39-38$ and $J=38-37$ transitions claimed by \citet{bel97}. They obtained a ($2\sigma$) column density upper limit of \cold{9.4}{10}, and concluded that the relative lack of HC$_{11}$N compared with the smaller cyanopolyynes may be due to the propensity of carbon chains longer than 10 C-atoms to form closed ring structures. 

TMC-1 also contains shorter carbon-chains with O and S terminations: C$_2$O, C$_3$O, C$_2$S, C$_3$S  \citep{ohi91,kai04}. Theoretical models of interstellar anion chemistry have predicted that longer carbon-chain oxides could be present in TMC-1.  
\citet{cor12} predicted detectable column densities for  C$_{6}$O, HC$_{6}$O, C$_{7}$O and HC$_{7}$O in dense interstellar clouds, where they were theorised to form as a result of associative electron detachment (AED) reactions between carbon chain anions (H)C$_n^-$ and atomic oxygen. Such reactions have been  observed in the laboratory by \citet{eic07}, and although these reactions were found to be rapid, the product branching ratios are currently unknown. {\citet{mcg17} considered the production of HC$_n$O radicals ($n=3-7$) based on a chemistry initiated by radiative association reactions between hydrocarbon cations and CO.} The importance of these reaction mechanisms can be assessed by searching for, and measuring the abundances of, the predicted carbon chain oxide molecules.
 
Aromatic compounds play a fundamental role in  prebiotic chemistry \citep[\eg][]{ehr06}.  However, despite strong observational evidence for their presence in high-excitation astronomical environments via infrared emission \citep{tie08},  there have been no firm spectroscopic identifications of specific polycyclic aromatic hydrocarbon (PAH) molecules in the interstellar medium so far. 
 Aromatic nitrogen heterocycles are of particular prebiotic importance and, in particular,  pyrimidine could be a possible precursor of  DNA nucleobases. Experiments have shown that energetic processing of interstellar ice analogues containing pyrimidine can form the nucleobases uracil, thymine and cytosine  \citep{mat13,nue14}, as well as the N-heterocycles quinoline and isoquinoline \citep{mat15}. As a permanent electric dipole moment results from the presence of a heteroatom in the aromatic structure,   rotational transitions are possible, however such heterocycles have not yet been detected, despite dedicated searches in star-forming regions and circumstellar envelopes (\citealt{kua03,cha05}; {\citealt{bru06}}). 
  
Experimental and theoretical studies indicate that gas-phase formation of pyrimidine and benzene could be viable in dark clouds  \citep{ber09,jon11}. The recent discovery of benzene and naphthalene in the volatile ices of comet 67P (Altwegg, K., private communication 2016)  provides impetus to renew searches for aromatic interstellar compounds. 

During preparation of this article, \citet{mcg17} published a detection of the carbon chain oxide HC$_5$O in TMC-1 using the GBT K-band receiver, as well as a tentative detection of HC$_7$O. The first detection of benzonitrile (C$_6$H$_5$CN) has also {been} submitted for publication \citep{mcg17b}
  
In this paper we report detections and column densities for HC$_7$O and two isotopologues of HC$_7$N. We also report non-detections of HC$_{11}$N, HC$_6$O, C$_6$O, C$_7$O, malononitrile,   quinoline, isoquinoline and pyrimidine, for which column density upper limits are presented.

\section{Observations}

Spectra were obtained towards the TMC-1 cyanopolyyne peak (at J2000 R.A. 04:41:42, decl. +25:41:27) using the Robert C. Byrd Green Bank Telescope (GBT) in October-November 2011. Observations were conducted using the central beam of the K-band focal plane array (KFPA) receiver, with a half-power beam width of $\approx37''$ on the sky. The GBT spectrometer was configured with a channel width of 12.2 kHz and $4\times50$~MHz spectral windows. As a result of the GBT's dynamic scheduling system, observations were performed under consistently good weather conditions, with system temperatures ($T_{sys}$) in the range 35-50~K and zenith opacities 0.02-0.05. Pointing was checked every 1-2 hours, and the mean pointing error over the 64 hours allocated to our program was $5.2''$. Spectra in the range 18.1-19.8~GHz were obtained using frequency switching, and the range 20.2-20.5~GHz was observed by position switching, using an uncontaminated reference position offset by $17'$ NE from the cyanopolyyne peak.

Data reduction was performed using the GBTIDL software, which included opacity correction, frequency-alignment of individual scans in the LSR frame and $T_{sys}$-weighted averaging. Reduced spectra were subsequently corrected for the main beam efficiency of 0.92. On-source integration times $\sim10$ hours per spectral setting resulted in RMS noise levels of 1-2~mK per channel. Targeted spectral line frequencies are shown in Table \ref{tab:spec}. We used line lists from the Cologne Database for Molecular Spectroscopy \citep{mul01} and Jet Propulsion Laboratory Molecular Spectroscopy web pages\footnote{https://spec.jpl.nasa.gov/}, augmented, where necessary, by our own extrapolations outside the laboratory data sets.  The energy level quantum numbering schemes adopted here differ for the various molecules and are detailed in the primary spectroscopic references. For HC$_7$O, we obtained additional GBT KFPA data from the recent study of \citet{mcg17}.

\begin{table}
\centering
\caption{Targeted line frequencies, transitions and upper-state energies 
\label{tab:spec}}
\begin{tabular}{lllrr}
\hline\hline
Rest Freq.    & Species                & Transition                             & $E_u$\\
(MHz)         &                        &                                        &  (K)  \\
\hline
 18106.283    & HC$_7$O                & $17-16e$                               &  7.5  \\
 18106.312    & HC$_7$O                & $16-15e$                               &  7.5  \\
 18107.873    & HC$_7$O                & $17-16f$                               &  7.5  \\
 18107.900    & HC$_7$O                & $16-15f$                               &  7.5  \\
 18334.033    & C$_7$O                 & $16-15$                                &  7.5  \\
 18596.764    & HC$_{11}$N             & $55-54$                                & 25.0  \\
 18638.616    & HC$_5$N                & $7-6$                                  &  3.6  \\
 18828.319    & HC$_6$O                & $12-11e$                               &  5.5  \\
 18828.336    & HC$_6$O                & $11-10e$                               &  5.5  \\
 18829.311    & HC$_6$O                & $12-11f$                               &  5.5  \\
 18829.327    & HC$_6$O                & $11-10f$                               &  5.5  \\
 18992.942    & C$_9$H$_7$N            & $10_{0,10}-9_{0,9}$, $F=10-9$          &  5.2  \\
 18992.990    & C$_9$H$_7$N            & $10_{0,10}-9_{0,9}$, $F= 9-8$          &  5.2  \\
 18993.002    & C$_9$H$_7$N            & $10_{0,10}-9_{0,9}$, $F=11-10$         &  5.2  \\
 19071.386    & C$_6$O                 & $11_{12}-10_{11}$                      &  5.9  \\
 19203.671    & HC$_7$O                & $18-17e$                               &  8.4  \\
 19203.699    & HC$_7$O                & $17-16e$                               &  8.4  \\
 19205.267    & HC$_7$O                & $18-17f$                               &  8.4  \\
 19205.294    & HC$_7$O                & $17-16f$                               &  8.4  \\
 19479.903    & C$_7$O                 & $17-16$                                &  8.4  \\
 19785.41     & CH$_2$(CN)$_2$         & $4_{1,3}-4_{0,4}$                      &  3.6  \\
 20278.94     & C$_4$H$_4$N$_2$        & $12_{9,3}-12_{8,4}$                    & 44.8  \\
 20287.345    & HC$_{11}$N             & $60-59$                                & 29.7  \\
 20292.487    & HC$_3$$^{13}$CC$_3$N   & $18-17$                                &  9.3  \\
 20294.271    & HC$_4$$^{13}$CC$_2$N   & $18-17$                                &  9.3  \\
 20303.946    & HC$_7$N                & $18-17$                                &  9.3  \\
 20465.490    & HC$_6$O                & $13-12e$                               &  6.5  \\
 20465.504    & HC$_6$O                & $12-11e$                               &  6.5  \\
 20466.662    & HC$_6$O                & $13-12f$                               &  6.5  \\
 20466.676    & HC$_6$O                & $12-11f$                               &  6.5  \\
 20529.076    & $i$-C$_9$H$_7$N      & $11_{ 1,11}-10_{1,10}$, $F=11-10$      &  6.1  \\
 20529.097    & $i$-C$_9$H$_7$N      & $11_{ 1,11}-10_{1,10}$, $F=10-9$       &  6.1  \\
 20529.113    & $i$-C$_9$H$_7$N      & $11_{ 1,11}-10_{1,10}$, $F=12-11$      &  6.1  \\
\hline
\end{tabular}
\parbox{\columnwidth}
{\footnotesize \vspace*{1mm} {\bf Notes.} Primary spectroscopic sources 
for molecular line frequencies: 
HC$_7$O and HC$_6$O -- \citet{hcno2005},
C$_7$O -- \citet{c7o1995},
HC$_{11}$N -- \citet{hc11n1996},
C$_9$H$_7$N and $i$-C$_9$H$_7$N -- \citet{qisoq2003},
C$_6$O -- \citet{c6o1995},
CH$_2$(CN)$_2$ -- \citet{malhyp1985},
C$_4$H$_4$N$_2$ -- \citet{prm1999},
HC$_5$N -- \citet{2004hc5n},
HC$_7$N and its isotopic species -- \citet{hc7n2000}. The CH$_2$(CN)$_2$ and C$_4$H$_4$N$_2$ lines are blends of many 
hyperfine components due to the presence of two nitrogen nuclei. }
\end{table}

%\newpage
\section{Results}

Column densities were obtained using spectral line models constructed for each species, based on Gaussian (thermally broadened) opacity profiles \citep[\eg][]{num00}. The cloud radial velocity (5.81~\kms) and Doppler FWHM (0.37~\kms) were obtained from a least-squares fit to the high signal-to-noise HC$_7$N $J=18-17$ line (shown in Figure \ref{fig:cyano}). The column density for each species was varied until the best fit to the observed spectral data was obtained; $1\sigma$ uncertainties were derived using Monte Carlo noise replication and re-fitting with 500 replications (as described by \citealt{cor13}).  For species with no detectable emission, we adopted an approach similar to \citet{loo16}, by constructing a spectral model (with fixed radial velocity and line FWHM), and increasing the column density from zero until a $\chi^2$ value of $2\sigma$ was reached. For species with multiple lines in different spectral windows, the fit was generated for all lines simultaneously, weighted by the RMS noise level in each window.

Rotational temperatures were taken from the literature where available: 5~K was adopted for HC$_5$N and 6~K for HC$_7$N based on \citet{bel98}, and 10~K for HC$_{11}$N based on \citet{bel97} and \citet{loo16}. Although the kinetic temperature of TMC-1 CP has been well established at 10~K \citep[\eg][]{pra97}, species with large dipole moments (such as the cyanopolyynes and related asymmetric species with 3-9 C atoms) undergo rapid rotational cooling that can cause their rotational excitation temperatures ($T_{rot}$) to fall away from thermodynamic equilibrium. As discussed by \citet{bel98}, the magnitude of this effect for the cyanopolyynes is primarily dependent on the rotational constant (inverse of the molecule length). Thus, for species containing a chain of 6 or 7 C-atoms, we adopted $T_{rot}=6$~K. For the remaining molecules, we adopted $T_{rot}=10$~K. Our calculated column densities, abundances and upper limits are given in Table \ref{tab:colds}. The TMC-1 H$_2$ column density of $N_{\rm H_2}=10^{22}$~cm$^{-1}$ was taken from \citet{cer87}.

\begin{table*}
\centering
\caption{TMC-1 (CP) molecular column density/abundance measurements and upper limits \label{tab:colds}}
\begin{tabular}{lllrr}
\hline\hline
Species&Name&$T_{rot}$~(K)&$N$~(cm$^{-2}$)&$N/N_{\rm H_2}$\\
\hline
HC$_{11}$N&Cyanodecapentayne&10&$<7.4\times10^{10}$&$<7.4\times10^{-12}$\\
HC$_{7}$N&Cyanohexatriyne&6&$1.36\times10^{13}$&$1.36\times10^{-9}$\\
HC$_3$$^{13}$CC$_3$N&$^{13}$C-Cyanohexatriyne&6&$(1.2\pm0.2)\times10^{11}$&$(1.2\pm0.2)\times10^{-11}$\\
HC$_4$$^{13}$CC$_2$N&$^{13}$C-Cyanohexatriyne&6&$(1.4\pm0.2)\times10^{11}$&$(1.4\pm0.2)\times10^{-11}$\\
HC$_{5}$N&Cyanobutadiyne&5&$5.3\times10^{13}$&$5.3\times10^{-9}$\\
C$_6$O&Hexacarbon monoxide&6&$<5.2\times10^{10}$&$<5.2\times10^{-12}$\\
C$_7$O&Heptacarbon monoxide&6&$<2.6\times10^{10}$&$<2.6\times10^{-12}$\\
HC$_6$O&1-oxo-hexa-1,3,5-triynyl&6&$<1.5\times10^{11}$&$<1.5\times10^{-11}$\\
HC$_7$O&1-oxo-hepta-2,4,6-triynyl&6&$(7.8\pm0.9)\times10^{11}$&$(7.8\pm0.9)\times10^{-11}$\\
CH$_2$(CN)$_2$&Malononitrile&10&$<7.3\times10^{10}$&$<7.3\times10^{-12}$\\
C$_4$H$_4$N$_2$&Pyrimidine&10&$<4.2\times10^{13}$&$<4.2\times10^{-9}$\\
C$_9$H$_7$N&Quinoline&10&$<6.2\times10^{11}$&$<6.2\times10^{-11}$\\
$i$-C$_9$H$_7$N&Isoquinoline&10&$<9.4\times10^{11}$&$<9.4\times10^{-11}$\\
\hline
\end{tabular}
\end{table*}

 %___________________________________________________________________________________________________
\subsection{Carbon chain oxides}

\begin{figure}
\centering
\includegraphics[width=\columnwidth]{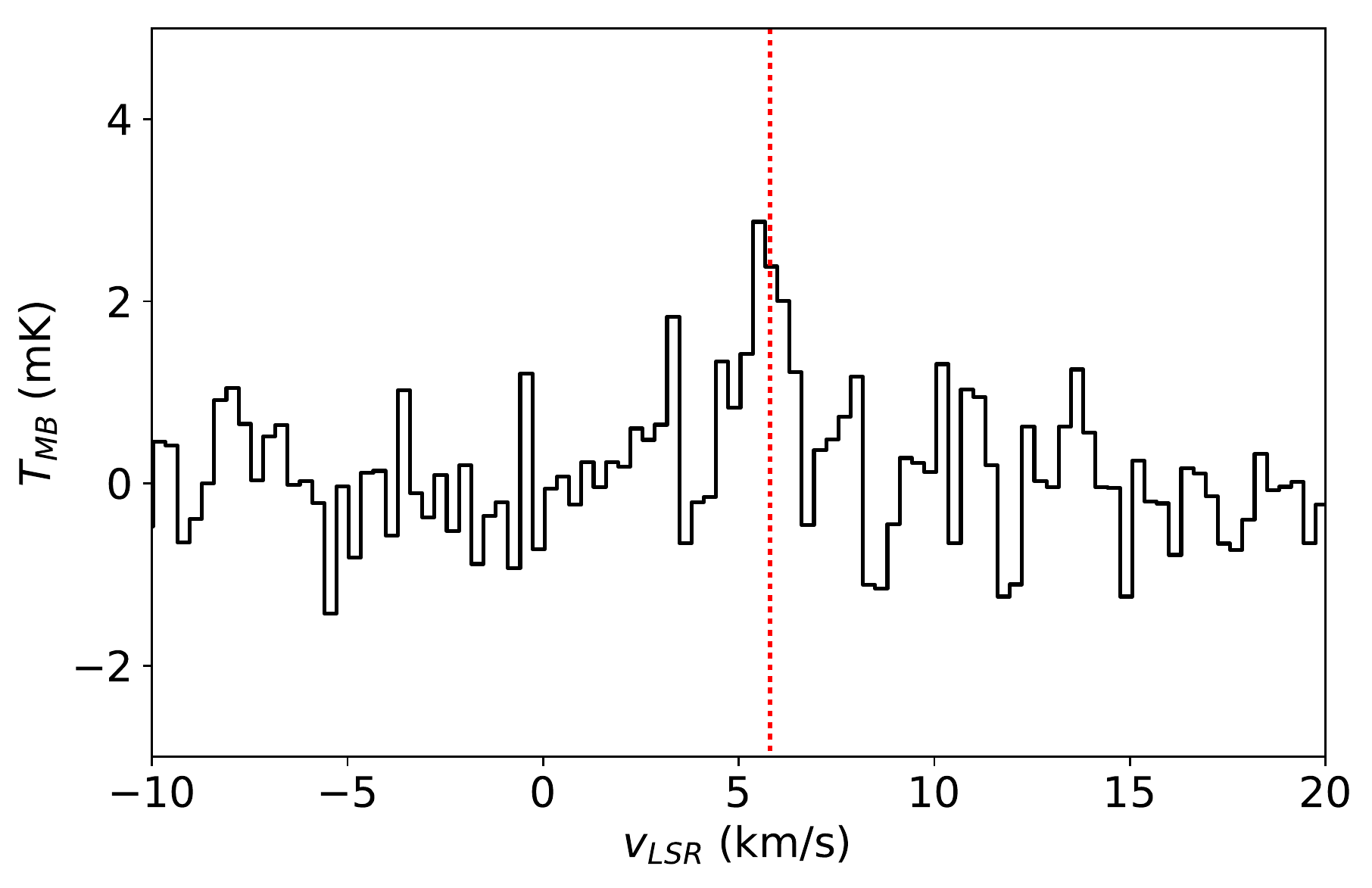}
\caption{Average (in velocity space) of the four HC$_7$O doublets in our observed GBT spectra, showing a clear detection at the expected position of 5.8~\kms (dotted line). \label{fig:meanhc7o}}
\end{figure}

To obtain improved sensitivity to the {cumulenone radical} species HC$_7$O, our observational data (covering the $F=17-16$ and $16-15$ $e$ and $f$ transitions) were combined with additional GBT data (covering the $F=18-17$ and $17-16$ $e$ and $f$ transitions) from the study of \citet{mcg17} for a total of eight HC$_7$O transitions, in the form of two pairs of hyperfine doublets.  The data were rebinned to a common frequency resolution and transformed to the (LSR) velocity scale. At the resolution of our observed spectra, the hyperfine doublet structure of each line was not well resolved, so the mean rest frequency was used for each doublet. The average spectrum for the four doublets is shown in Figure \ref{fig:meanhc7o}, with a well-defined, statistically significant peak at the TMC-1 systemic velocity of 5.8~\kms\ (marked with a vertical dotted line). The integrated line intensity is $3.8\pm0.39$~mK\,\kms, which corresponds to a $9.5\sigma$ detection.

\begin{figure}
\centering
\includegraphics[width=\columnwidth]{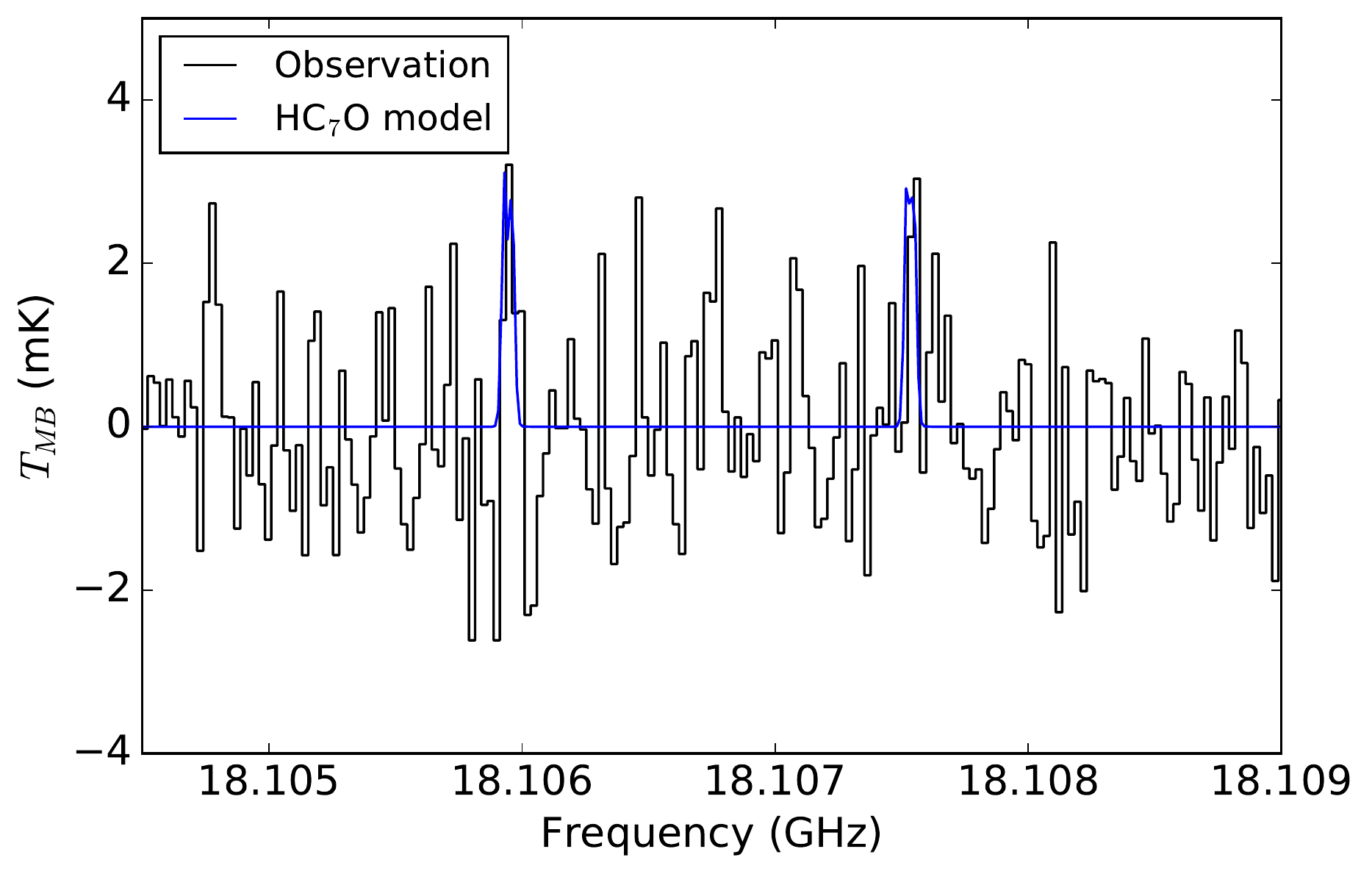}\\
\includegraphics[width=\columnwidth]{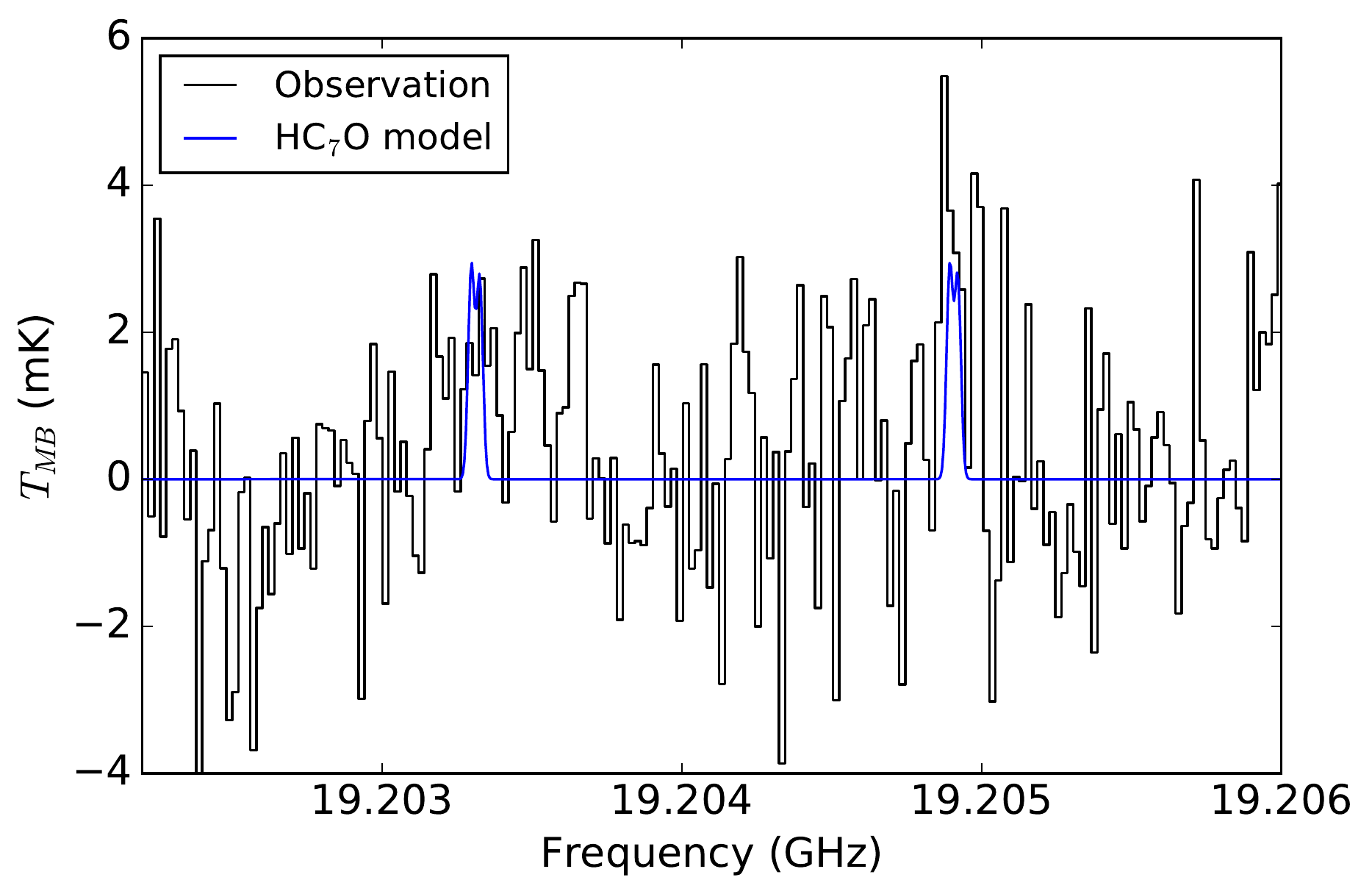}
\caption{GBT spectra of the regions surrounding the two sets of observed HC$_7$O transitions (top panel: $17-16e$, $16-15e$, $17-16f$, $16-15f$; bottom panel: $18-17e$, $17-16e$, $18-17f$, $17-16f$). The same (best-fitting) HC$_7$O model is overlaid on each. Taken separately, these spectral features are of insufficient intensity to claim a detection, but taken together, they imply a detection of this molecule at the $9.5\sigma$ confidence level. \label{fig:hc7o}}
\end{figure}

A spectral line model was constructed to fit all eight HC$_7$O hyperfine components, shown in Figure \ref{fig:hc7o}. Allowing both the HC$_7$O column density ($N$) and radial velocity ($v$) to vary, the best-fitting model results were $N({\rm HC_7O})=(7.8\pm0.9)\times10^{11}$~cm$^{-2}$, and $v=5.83\pm0.05$~\kms.

{Three possible formation mechanisms have been proposed for these carbon chain oxides.  The observations reported here were originally motivated by the theoretical models of}   \citet{cor12} who considered the formation of C$_6$O,  C$_7$O,  HC$_6$O and HC$_7$O as a result of reactions between O atoms and carbon chain anions, as measured in the laboratory by \citealt{eic07}. For reactions of oxygen with the polyyne anions C$_n$H$^-$ $n=2-7)$, the most likely product channels are:

\be
{\rm C}_n{\rm H}^- + {\rm O} \longrightarrow {\rm HC}_n{\rm O} + e^-
\ee

\be
{\rm C}_n{\rm H}^- + {\rm O} \longrightarrow {\rm C}_n{\rm O}^- + {\rm H}
\ee

\be
{\rm C}_n{\rm H}^- + {\rm O} \longrightarrow {\rm C}_{n-1}{\rm H}^- + {\rm CO} 
\ee

{The chemical model for TMC-1 presented by \citet{cor12} predicted a maximum HC$_7$O column density of $\approx 1.5\times10^{13}$~cm$^{-2}$ at a chemical age of $t \approx 1\times10^{5}$ years, as well as potentially detectable abundances of the other three molecules.  This calculated maximum HC$_7$O  abundance is about a factor of 19 higher than the observed value reported here, and the modeled abundances of C$_6$O,  C$_7$O and  HC$_6$O are also significantly higher than our derived upper limits.
  However,  comparison of the observed  HC$_7$O  abundance ($\approx 8\times10^{-11}$) with the TMC-1 model  (Figure 4 of \citealt{cor12}) indicates that it can be reproduced at $t \approx 3\times10^{5}$ years, at which point the C$_6$O and C$_7$O abundances are also within the upper limits reported here. Although at $t \approx 3\times10^{5}$ years  the HC$_6$O  abundance is comparable to that of HC$_7$O,  in conflict with our observed abundance limits,   it should be noted that the \citet{cor12} model also over-predicted the C$_6$H$^-$ abundance (by a factor of 6.7, compared with the observed value from \citealt{bru07}). If the modeled C$_6$H$^-$ abundance was less by a similar factor, then the predicted HC$_6$O abundance  would drop to just below our observed upper limit as a result of the close chemical relationship between these species.

\citet{mcg17} considered the chemistry of the HC$_n$O radicals ($n=3-7$) based on an extension of the reactions suggested by \citet{ada89}. In this case the cumulenone radicals are produced in radiative association reactions between hydrocarbon cations and CO: 
  
\be
{\rm C}_{n-1}{\rm H_2}^+ + {\rm CO} \longrightarrow {\rm H_2C}_n{\rm O^+} + \nu
\ee

\be
{\rm C}_{n-1}{\rm H_3}^+ + {\rm CO} \longrightarrow {\rm H_3C}_n{\rm O^+} + \nu
\ee

 followed by electron dissociative recombination of the ions.  In a model for TMC-1, \citet{mcg17} found that their observed  HC$_5$O column density and their HC$_7$O  upper limit were best reproduced at $t \approx 2\times10^{5}$ years. However,  at this time HC$_4$O was calculated to be about 37 times more abundant than observed, and HC$_6$O was a factor of $3\times10^{-3}$ below the observed upper limit.
 
  It has been suggested that the presence  of CCO and CCCO in TMC-1, as well as  perhaps other carbon chain oxides,  could originate in the injection  of surface-formed cumulenone molecules from dust grains \citep{mar00}. On cold dust grains,  carbon atom additions, starting from HCO, could possibly lead to  long cumulenone HC$_n$O  radicals  that would form cumulenone molecules, H$_2$C$_n$O, following an H atom addition (cf. \citealt{cha97}).
   Following desorption,  the cumulenone molecules would be protonated and the product ions subject to electron dissociative recombination reactions,  leading to  C$_n$O and   HC$_n$O  radicals, as well as  H$_2$C$_n$O molecules, all being present  in the gas.   Thus, in this case, the HC$_n$O  radicals would have both a grain-surface and a gas-phase origin.  However, the non-detection of the propynonyl  radical (HC$_3$O)   in TMC-1 \citep{mcg17}, and consistently negative results from searches for propadienone (H$_2$C$_3$O)  in  molecular clouds,  including TMC-1 \citep{irv88, bro92, loi16}, tend to strongly disfavor this scenario. 
    
  Theoretical models for carbon chain oxide chemistry will require laboratory measurements  of the rates and branching ratios of key formation reactions. Generally, the product distributions of electron dissociative reactions involving $ {\rm H_mC}_n{\rm O^+}  $ ions are important and, 
 for anion chemistry, an assessment of  the efficiency of the associative electron detachment (AED) channel (Reaction 1) would be very informative.  
    Future deep searches in TMC-1 and elsewhere for these molecules will permit the  elucidation of their exact formation pathways; for example, a detection of HC$_6$O would help distinguish between the anionic and cationic formation mechanisms.}

%___________________________________________________________________________________________________
\subsection{Cyanopolyynes}

\subsubsection{HC$_{11}$N confirmed non-detection}

\begin{figure}
\centering
\includegraphics[width=\columnwidth]{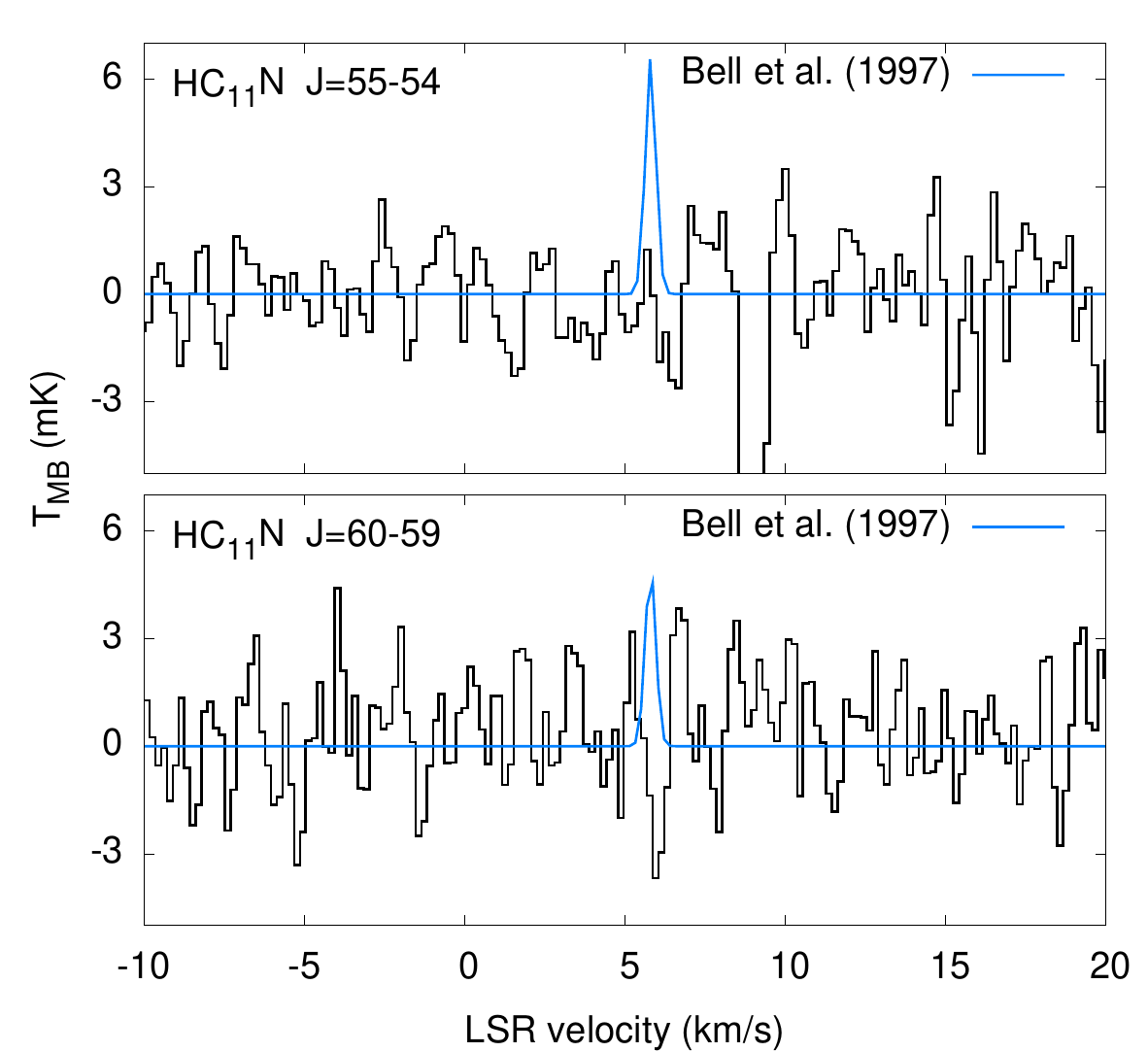}
\caption{GBT K-band spectra surrounding two transitions of HC$_{11}$N. Predicted spectra based on the column density reported by \citet{bel97} are overlaid in blue. \label{fig:hc11n}}
\end{figure}

Our GBT K-band spectra support the results of \citet{loo16}: we searched at the frequencies of the higher-energy $J=55-54$ and $J=60-59$ transitions but found no evidence for any HC$_{11}$N emission. Our spectra are shown in Figure \ref{fig:hc11n} with a model based on the HC$_7$N line FWHM and radial velocity, with the temperature and column density of \citet{bel97} overlaid. Our spectral line modeling implies a column density upper limit of \cold{7.4}{10} (at 95\% confidence), which is somewhat lower than the upper limit obtained by \citet{loo16}, and almost four times less than the value of \cold{2.8}{11} claimed by \citet{bel97}.

We performed our search for HC$_{11}$N at the same coordinates as \citet{loo16}, so our combined results are consistent with a lack of HC$_{11}$N at the TMC-1 cyanopolyyne peak. It should be noted, however, that the detection claimed by \citet{bel97} was based on observations using a significantly larger ($2.4'$) telescope beam, offset from the nominal cyanopolyyne peak by about 1~s in RA and $9''$ in declination. It is therefore conceivable that their larger, slightly offset telescope beam was sensitive to HC$_{11}$N emission that was missed by the present study and that of \citet{loo16}. The previous detection of HC$_{11}$N in TMC-1 therefore cannot be completely refuted until additional observations are performed, ideally covering the same area on the sky as those made by \citet{bel97} using the NRAO 140-foot telescope.

\vspace*{5mm}
\subsubsection{New HC$_{7}$N isotopologues}
  
\begin{figure}
\centering
\includegraphics[width=\columnwidth]{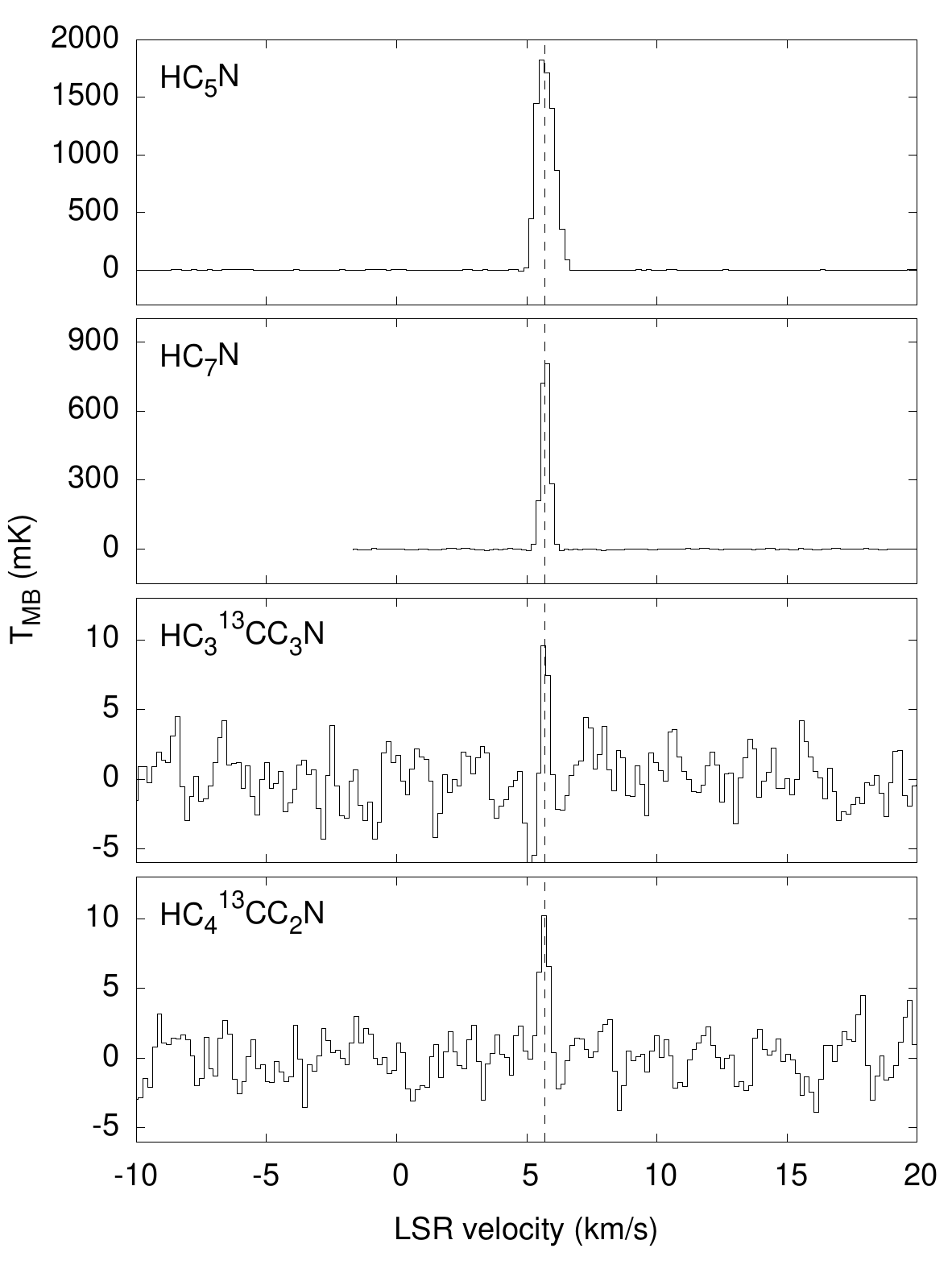}
\caption{GBT K-band spectra showing detections of HC$_5$N, HC$_7$N, and its $^{13}$C-substituted isotopologues HC$_3$$^{13}$CC$_3$N and HC$_4$$^{13}$CC$_2$N. \label{fig:cyano}} \end{figure}

 The spectra shown in Figure \ref{fig:cyano} reveal the presence of the $J = 18-17$ lines of HC$_3$$^{13}$CC$_3$N and HC$_4$$^{13}$CC$_2$N at the 6-7$\sigma$ level. This represents the first reported detection of individual isotopologues of HC$_7$N in any environment we are aware of. The only prior observation of $^{13}$C-substituted HC$_7$N was in TMC-1 by \citet{lan07} using the GBT, who stacked the spectra of various HC$_7$N isotopologues (with $^{13}$C at different positions in the carbon chain), to obtain a mean $^{12}$C/$^{13}$C ratio of $87^{+37}_{-19}$ for this molecule.   We derive abundance ratios for the specific isotopologues HC$_7$N/HC$_3$$^{13}$CC$_3$N = $110\pm16$ and HC$_7$N/HC$_4$$^{13}$CC$_2$N = $96\pm11$. These values are notably high, and indicate depletion of $^{13}$C relative to the local interstellar elemental $^{12}$C/$^{13}$C ratio of 60-70 \citep{luc98,mil05}. Depleted $^{12}$C/$^{13}$C ratios have previously been observed in TMC-1 for the closely related molecules HC$_3$N and HC$_5$N, as well as CCH and CCS \citep{sak13}.  Modeling efforts are in progress to understand this isotopic depletion, which seems to be a common characteristic of carbon chains in cold interstellar gas \citep{yos15,ara16}. 
 
Based on our new results, the depletion of $^{13}$C in HC$_7$N (resulting in a higher $^{12}$C/$^{13}$C ratio), seems to be stronger than for the shorter cyanopolyynes. For HC$_7$N, our mean $^{12}$C/$^{13}$C ratio is $103\pm6$, compared with the mean value of $94\pm6$ for HC$_5$N \citep{tan16} and $70\pm5$ for HC$_3$N \citep{tak98}. However, it should be noted that previous studies have found the ratio to be significantly smaller for the carbon atom in the CN end-group than for the other C atoms along the chain \citep{tan16}. Excluding the CN end-group, the mean ratios are $97\pm7$ for HC$_5$N and $77\pm7$ for HC$_3$N. Additional observations will be required to determine if this is also the case for HC$_7$N.
 
The current understanding regarding the origin of $^{13}$C depletion in carbon chain-bearing species is that it likely occurs due to gas-phase isotope exchange reactions, whereby $^{13}$C is preferentially incorporated into other molecules. For instance, it is known that the $^{12}$C/$^{13}$C ratio in (atomic) C$^{+}$ becomes depleted due to the incorporation of the heavier isotope into CO, through the exchange of $^{12}$C and $^{13}$C in the reaction between $^{13}$C$^+$ and $^{12}$CO \citep{lan84}. It has thus been hypothesized that as gas-phase C$^{+}$ becomes depleted in $^{13}$C, any molecules whose chemistry is closely linked with C$^{+}$ should also show isotopic depletion \citep{sak13}.

It has been suggested that the analogous exchange reaction of $^{13}$C$^+$ with $^{12}$CN could produce the observed (relative) $^{13}$C enrichment of the CN group in cyanopolyynes \citep{tak98,ara16}, but the viability of this mechanism has yet to be conclusively demonstrated in dense molecular clouds. Based on the relative similarities of the $^{12}$C/$^{13}$C ratios in the different C atoms of HC$_5$N in TMC-1, \citet{tan16} deduced that the main route to synthesizing the observed HC$_5$N is via hydrocarbon ion (plus N-atom) chemistry. The trend for lower $^{13}$C fractions in progressively larger cyanopolyynes is therefore indicative of a diminishing $^{13}$C content in hydrocarbon ions of increasing size.

\subsection{Nitrogen heterocycles}

We searched for emission from multiple lines from quinoline (C$_9$H$_7$N), isoquinoline ($i$-C$_9$H$_7$) and pyrimidine (C$_4$H$_4$N$_2$) in TMC-1 and obtained strict (95\% confidence) upper limits on their column densities of $6.2\times10^{11}$~cm$^{-2}$ and $9.4\times10^{11}$~cm$^{-2}$ for quinoline and isoquinoline, respectively, and $4.2\times10^{13}$~cm$^{-2}$ for pyrimidine. [We also obtained an upper limit of $7.3\times10^{10}$~cm$^{-2}$ for malononitrile (CH$_2$(CN)$_2$).]

\begin{figure}
\centering
\includegraphics[width=\columnwidth]{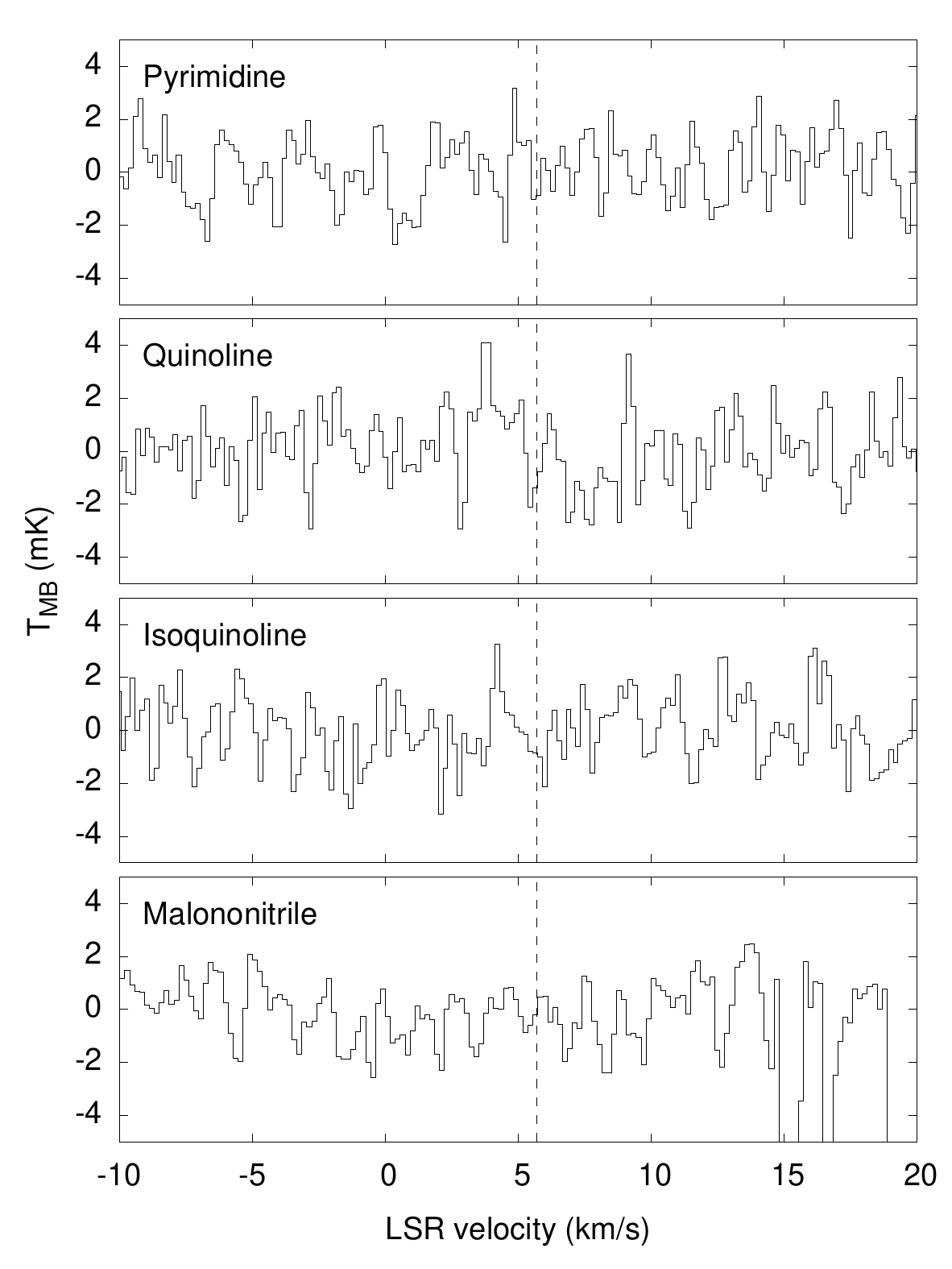}
\caption{GBT K-band spectra showing non-detections of pyrimidine (C$_4$H$_4$N$_2$), quinoline (C$_9$H$_7$N), isoquinoline ($i$-C$_9$H$_7$) and malononitrile (CH$_2$(CN)$_2$). \label{fig:bio}} \end{figure}

\vspace*{5mm}
\section{Conclusions }

We have performed deep searches for large organic molecules in TMC-1 with the GBT. Our data, when combined with that of \citet{mcg17}, allows us to confirm the previous tentative detection of HC$_7$O. The measured HC$_7$O  column density and the derived upper limits for 
 C$_6$O,  HC$_6$O and C$_7$O  place constraints on chemical models in which these molecules form through reactions involving {negative and/or positively-charged hydrocarbon ions}. 
 We also confirm the previously reported non-detection of HC$_{11}$N at the cyanopolyyne peak by \citet{loo16}. 
  We find that the detected  $^{13}$C isotopologues of HC$_{7}$N are depleted in $^{13}$C. Reproduction of these $^{12}$C/$^{13}$C ratios, as well as those measured in  shorter carbon chain molecules (CCH, CCS, HC$_{3}$N and HC$_{5}$N),   present a challenge for
   models of interstellar isotopic fractionation.
     Finally, our non-detections of  nitrogen heterocycles in a cold molecular cloud complement previous non-detections of 
   aromatic molecules in star-forming regions and in the envelopes of evolved stars. This raises the question: why have no aromatic compounds yet been detected by astronomical mm/submm spectroscopy?   Most previous searches  have involved targeting molecules in which the   heteroatom, specifically N, is incorporated  in an aromatic ring. 
   While our   derived abundance of pyrimidine is not particularly restrictive with regard to its possible presence in TMC-1,    the recent detection of benzonitrile in TMC-1 by  \citet{mcg17b} and of toluene in comet 67P/Churyumov-Gerasimenko \citet{alt17} suggests that interstellar chemistry may favor the presence  of heteroatoms as functional side-groups, rather than within the ring structure.
     Future deep searches in TMC-1 with the GBT  should provide further insights into the inventory of the heaviest interstellar molecules.

%______________________________________________________________________________________________________

%______________________________________________________________________________________________________

\acknowledgements
This work was supported by the Goddard Center for Astrobiology and NASA's Origins of Solar Systems and Exobiology programs.

%%%%%%%%%%%%%%%%%%%%%%%%%%%%%%%%%%%%%%%%%%%%%%%%%%

%%%%%%%%%%%%%%%%%%%% REFERENCES %%%%%%%%%%%%%%%%%%

% The best way to enter references is to use BibTeX:

\bibliographystyle{aa}
\bibliography{refs} % if your bibtex file is called example.bib

%%%%%%%%%%%%%%%%%%%%%%%%%%%%%%%%%%%%%%%%%%%%%%%%%%

%%%%%%%%%%%%%%%%% APPENDICES %%%%%%%%%%%%%%%%%%%%%

%%%%%%%%%%%%%%%%%%%%%%%%%%%%%%%%%%%%%%%%%%%%%%%%%%

\end{document}